\documentclass[conference]{IEEEtran}
\IEEEoverridecommandlockouts
\usepackage[utf8]{inputenc}
\usepackage{times}
\usepackage{color}
\usepackage{graphicx}
\usepackage{subcaption}
\usepackage{pgf}
\usepackage{tikz}
\usetikzlibrary{arrows,shapes,fit,automata,positioning,decorations,calc} 
\usetikzlibrary{spy,backgrounds}

\usepackage{amsmath}
\usepackage{xspace}
\usepackage{tabularx}
\usepackage{colortbl}
\usepackage{nicefrac}
\usepackage{stackengine}
\usepackage{setspace}
\usepackage{todo}
\usepackage{picins}
\usepackage{wrapfig}
\usepackage{tabularx}

\usepackage{amsfonts}
\usepackage{mathrsfs}
\usepackage{marvosym}
\usepackage[T1]{fontenc}
\usepackage{pifont}
\usepackage{multirow}

\newcommand{\ignore}[1]{{}}

\newcolumntype{L}[1]{>{\raggedright\let\newline\\\arraybackslash\hspace{0pt}}m{#1}}
\newcolumntype{C}[1]{>{\centering\let\newline\\\arraybackslash\hspace{0pt}}m{#1}}
\newcolumntype{R}[1]{>{\raggedleft\let\newline\\\arraybackslash\hspace{0pt}}m{#1}}

\newcommand{\false}{\ensuremath{\mathsf{F}}}

\newcolumntype{L}[1]{>{\raggedright\let\newline\\\arraybackslash\hspace{0pt}}m{#1}}
\newcolumntype{C}[1]{>{\centering\let\newline\\\arraybackslash\hspace{0pt}}m{#1}}
\newcolumntype{R}[1]{>{\raggedleft\let\newline\\\arraybackslash\hspace{0pt}}m{#1}}

\newcommand{\squishlist}{
 \begin{list}{$\bullet$}
  { \setlength{\itemsep}{0pt}
     \setlength{\parsep}{1pt}
     \setlength{\topsep}{1pt}
     \setlength{\partopsep}{0pt}
     \setlength{\leftmargin}{0.9em}
     \setlength{\labelwidth}{1.5em}
     \setlength{\labelsep}{0.4em} } }
\newcommand{\squishend}{
  \end{list}  }

{\par\medskip\noindent\minipage{\linewidth}}
{\endminipage\par\medskip}

\definecolor{graphFirst}{RGB}{20,200,100} 
\definecolor{red1}{RGB}{47,47,211} 
\definecolor{graphSecond}{RGB}{1,1,1} 
\definecolor{graphThird}{RGB}{245,24,24} 
\definecolor{graphFourth}{RGB}{56,142,60} 
\definecolor{graphFifth}{RGB}{81,45,168} 
\definecolor{graphSixth}{RGB}{69,90,100} 
\definecolor{graphSeventh}{RGB}{251,192,45} 

\definecolor{backgroundFirst}{RGB}{129,212,250} 
\definecolor{backgroundSecond}{RGB}{154,154,239} 
\definecolor{backgroundThird}{RGB}{255,150,100} 
\definecolor{backgroundFourth}{RGB}{165,214,167} 
\definecolor{backgroundFifth}{RGB}{179,157,219} 
\definecolor{backgroundSixth}{RGB}{176,190,197} 
\definecolor{backgroundSeventh}{RGB}{255,245,157} 
\newtheorem{example}{Example}

\begin{document}
	\title{Runtime Monitoring and Statistical Approaches for Correlation Analysis of ECG and PPG
}

\author{
	\IEEEauthorblockN{Abhinandan Panda}
	\IEEEauthorblockA{\textit{School of Electrical Sciences} \\
		\textit{IIT Bhubaneswar}\\
		Bhubaneswar, India \\
		Email: ap53@iitbbs.ac.in}
	\and
	\IEEEauthorblockN{Srinivas Pinisetty}
	\IEEEauthorblockA{\textit{School of Electrical Sciences} \\
		\textit{IIT Bhubaneswar}\\
		Bhubaneswar, India \\
		Email: spinisetty@iitbbs.ac.in}
	\and
	\IEEEauthorblockN{Partha Roop}
	\IEEEauthorblockA{\textit{Dept. of Electr. \& Comput. Eng.} \\
		\textit{University of Auckland}\\
		Auckland, New Zealand\\
		Email: p.roop@aucklanduni.ac.nz}
}

\maketitle

\begin{abstract}
Biophysical signals such as Electrocardiogram (ECG) and Photoplethysmogram (PPG) are
key to the sensing of vital parameters for wellbeing. Coincidentally, ECG and PPG are signals, which provide a "different window" into the same phenomena, namely the cardiac cycle. While they are used separately, there are no studies regarding the exact correction of the different ECG and PPG events. Such correlation would be helpful in many fronts such as sensor fusion for improved accuracy using cheaper sensors and attack detection and mitigation methods using multiple signals to enhance the robustness, for example. Considering this, we present the first approach in formally establishing the key relationships between ECG and PPG signals. We combine formal run-time monitoring with statistical analysis and regression analysis for our results.

\end{abstract}


\section{Introduction}
\label{sec:intro}

Biophysical signals such as Electrocardiogram (ECG) and Photoplethysmogram (PPG) carry vital information about human physiology and are key to the sensing of vital parameters for well-being. Coincidentally, both ECG and PPG represent the same phenomena, i.e. cardiac activity. Further, the availability of wearable ECG, PPG sensors facilitated the analysis of health parameters such as heart health, blood pressure (BP), oxygen saturation levels, blood glucose etc., using ECG and PPG \cite{swapna2020diabetes,li2018design,murat2020application}  \cite{mousavi2019blood,karimipour2009diabetic,chowdhury2016real}.
 
 While both signals are used separately, there are no studies regarding the exact correlation of the different ECG and PPG events and intervals. Understanding the correlation between the two will provide better insight for understanding the health parameters. Further, such correlation would be helpful for improved accuracy using cheaper sensors, attack detection and mitigation.
 
 In this regard, no formal detailed studies exist that provide detailed correlation. The work in \cite{banerjee2014photoecg} proposes machine learning approaches to estimate ECG parameters using PPG features, but without any detailed correlation. In \cite{wu2016new}, in order to analyse the BP using ECG, the authors have correlated ECG intervals with PPG intervals but without any justification. Also, there is no established method nor formal validation of the correlations between the two signals.

Considering this, we present the first approach in formally establishing the key relationships between ECG and PPG signals. 

\emph{Overview of the proposed approach:} We study the simultaneous recordings of ECG and PPG and extract features such as various events and intervals of both the signals. Based on manual analysis, we propose correlation between events and intervals of ECG-PPG. We perform statistical analysis for validating correlations among ECG-PPG intervals. For strengthening the proposed correlation, formal runtime monitoring (RV) is used as an additional approach. Regression analysis is used to validate the linear relationship and lag time between ECG and PPG events.

This paper is organized as follows: In Section \ref{sec:prelim}, we briefly present the basics of ECG and PPG and in Section \ref{sec:ecgPPG}, we propose  the correlation of ECG and PPG. In Section \ref{sec:validation}, we discuss the validation of ECG-PPG correlation using formal RV monitoring, statistical analysis and regression analysis. Finally conclusions are drawn in Section \ref{sec:conclusion}.


\section{ECG-PPG Prelimnaries}
\label{sec:prelim}

A cardiac cycle represents human heart activity from one heartbeat to another. It consists of two periods: one during which the heart muscle relaxes and refills with blood, called diastole, following a period of heart contraction and pumping of blood, called systole. Both ECG and PPG present systole and diastole of a cardiac cycle.   

\begin{figure}[hbt!]
	\centering

	{
		\includegraphics[width=0.8\linewidth, height=\textheight,keepaspectratio]{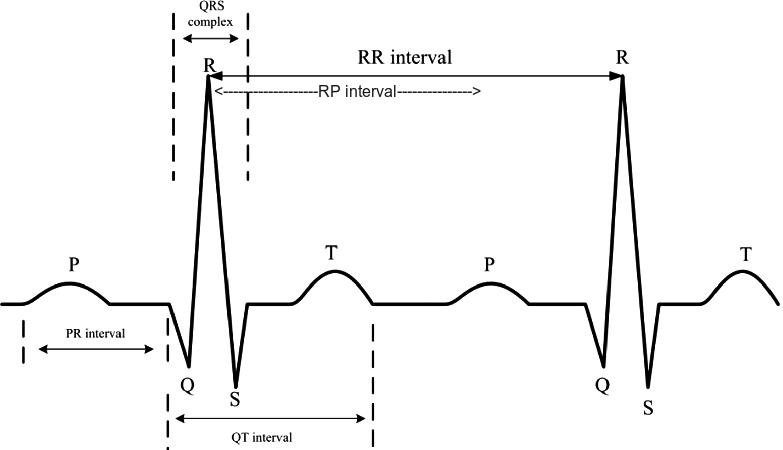}
		\vspace*{-2mm}
		\caption{A typical ECG signal.}
		\label{fig:ECGSignal}
	}

\end{figure}

\emph{ECG:} The Electrocardiogram (ECG) is the waveform produced by the heart showing the electrical activity of the heart over a period of time. A typical ECG signal is illustrated in Figure \ref{fig:ECGSignal}. The P-wave symbolizes atrial depolarization (indicating an atrial event has occurred). The PR interval is the time interval between the beginning of the P-wave and the beginning of the Q-wave. It shows the time taken by the electrical impulses to travel between the atria and ventricles. The QRS complex denotes the ventricular depolarisation (indicating ventricular event has occurred), atrial repolarization also happens during this interval. The T-wave represents ventricular depolarisation. The RR interval indicates the time interval between two QRS complexes.

\begin{figure}[hbt!]
	\centering

	{
		\includegraphics[width=0.7\linewidth, height=\textheight,keepaspectratio]{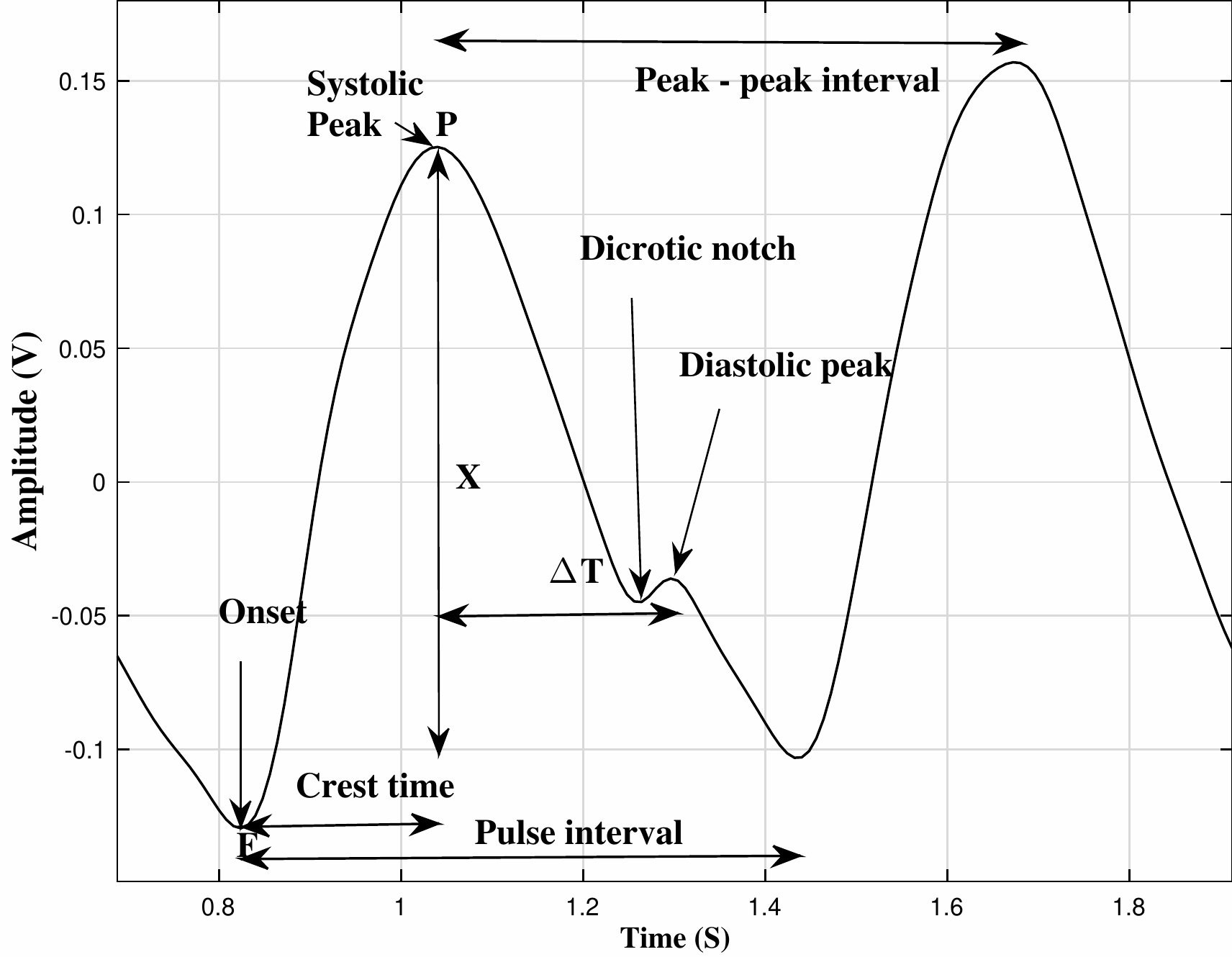}
		\vspace*{-2mm}
		\caption{A typical PPG signal.}
		\label{fig:PPGSignal}
	}

\end{figure}

\emph{PPG:} Photoplethysmogram (PPG) represents  the variations in blood volume or blood flow in the body which goes from the heart to the fingertips and toes through the blood vessels within a cardiac cycle \cite{awad2007relationship}. The various events and intervals of a PPG are shown in Figure \ref{fig:PPGSignal}.
	
The time interval between beginning and end of the PPG signal, shown in Figure \ref{fig:PPGSignal}, is known as the \emph{pulse interval}. The distance between two consecutive systolic peaks is referred to as \emph{peak-to-peak} interval. The time interval between the systolic and diastolic peak \emph{($\Delta$T)} is related to the time taken for the pulse wave to propagate from the heart to the periphery and back.  \emph{Crest} or \emph{systole period} is the time taken by PPG from the onset to its systolic peak. Similarly, the time taken by the signal in the diastole phase (systolic peak to next onset) is known as diastole period.


\vspace{0.5em}
\section{Manual analysis of ECG-PPG}
\label{sec:ecgPPG}
In this section, we present the observations regarding ECG-PPG correlation considering manual analysis of ECG-PPG events and intervals.

\begin{figure}[htp!]
	
	\centering

	{
		\includegraphics[width=0.7\linewidth, height=\textheight,keepaspectratio]{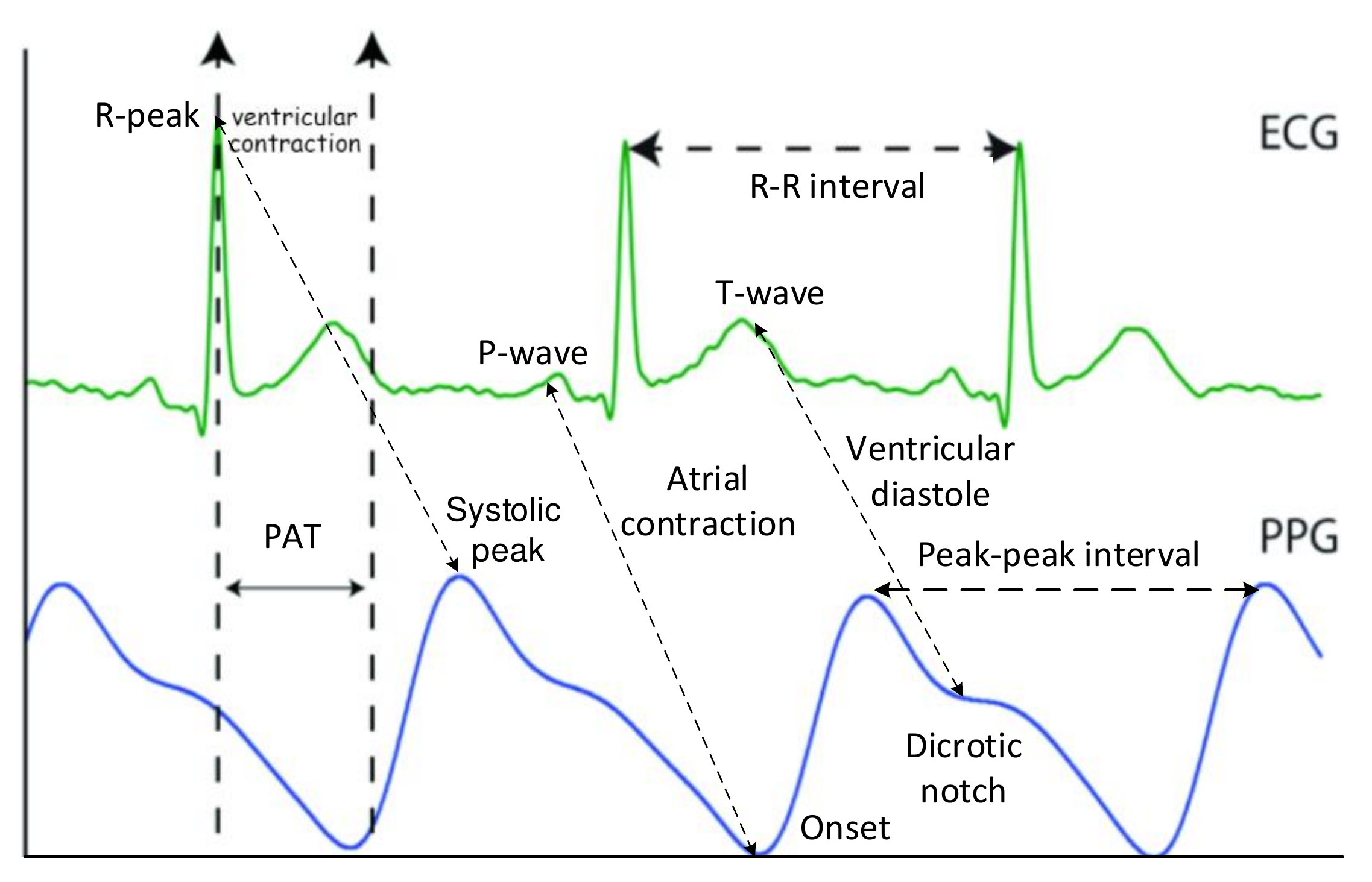}
		\caption{Simultaneous record of ECG and PPG showing cardiac activities.}
		\label{fig:ECGPPG}
	}
		
\end{figure}

We studied the simultaneous recordings of ECG-PPG available in the BIDMC database under physionet \cite{goldberger2000physiobank}. The database contains 53 recordings, each of 8 minutes duration. The signals were pre-processed for removal of noise. The signals are processed using  Neurokit2 tool\footnote{Neurokit is a Python toolbox for physiological signal processing. It implements signal processing algorithms such as Pan-Tomkin's algorithm, discrete wavelength trasform (DWT) etc. in python.}  \cite{Neurokit2} to extract events  such as P-peak, R-peak, Q-peak and T-peak of ECG and onset and systolic peak of PPG. We thereby computed ECG intervals such as PR, QR, RR, RP, RT, QT and systole and diastole period of PPG considering approximately 7586 samples of the dataset.

It has been studied that the systolic peak of a PPG signal corresponds to the R-peak of an ECG signal as both reflect ventricular contraction \cite{elgendi2012analysis}. Considering this as the starting point of our analysis, we detect other correlated events and intervals from manual analysis of ECG-PPG. It is observed that PR intervals of ECG are in the similar range of systole periods of PPG. Since, R-peak of ECG corresponds to the systolic peak of PPG, it may be assumed that the P-wave of ECG may correspond to the onset of PPG. It is observed that RR intervals of ECG are similar to peak-to-peak intervals of PPG. Similarly, the RP intervals of ECG are found to be in the similar range of diastole periods of PPG. During the analysis it is found that there is a delay between  the correlated events such as R-peak of ECG and systolic peak of PPG and P-peak of ECG and onset of PPG in the range 600-700 ms. The summary of observations from the manual analysis of ECG-PPG  are as follows.

\squishlist
	\item 
	$Obs_1$: The RR interval of ECG may be correlated to the peak-to-peak interval of PPG.
	
\item 
	$Obs_2$: The P-peak of ECG may correspond to onset of PPG signal.
	
	\item
	$Obs_3$: The PR interval of ECG may be correlated to the systole period of PPG.
	
	\item 
	$Obs_4$: The RP interval of ECG may be correlated to the diastole period of PPG.
	
	\item $Obs_5$: The onset of PPG may lag the P-peak of ECG by 600 - 700 ms (pulse arrival time (PAT)).
	\item $Obs_6$: The systolic peak of PPG may lag  the R-peak of ECG by 600 - 700 ms (pulse arrival time (PAT)).
\squishend

The summary of correlated ECG-PPG events and intervals (observations) are presented in Table \ref{tab:summary}. Figure \ref{fig:ECGPPG} presents cardiac activities along with corresponding ECG-PPG events. 

\begin{table}[]
	\centering
	
	\resizebox{0.8\columnwidth}{!}{%
		\begin{tabular}{|c|c|}
			\hline
			
			\textbf{\begin{tabular}[c]{@{}c@{}}ECG \\ Events \& Timers\end{tabular}} &
			\textbf{\begin{tabular}[c]{@{}c@{}}PPG \\ Events \& Timers\end{tabular}} \\ \hline
			 P-peak                & Onset                                            \\ \hline
			 R-peak                 & Systolic peak                                    \\ \hline
			 PR interval           & Systole period                     \\ \hline
			 RP interval           & Diastole period                     \\ \hline
			 RR interval  & peak-to-peak interval  \\ \hline
		\end{tabular}}
	\vspace{-2mm}
	\caption{Summary of correlated ECG, PPG events and timers.}
		\label{tab:summary}
		
\end{table}

\section{Validation of Proposed ECG-PPG Correlation}
\label{sec:validation}

In this section, the proposed correlation in Section \ref{sec:ecgPPG} is validated using a runtime monitoring framework, statistical analysis and regression analysis.

\begin{figure}[htb!]
	\begin{subfigure}{.25\textwidth}
		\centering
		\includegraphics[width=.8\linewidth]{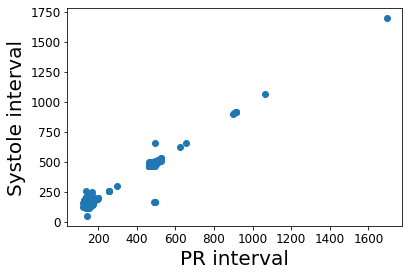}
		\caption{Scatter plot of PPG systole \\ period and ECG PR interval}
		\label{fig:sfig1}
	\end{subfigure}%
	\begin{subfigure}{.25\textwidth}
		\centering
		\includegraphics[width=.8\linewidth]{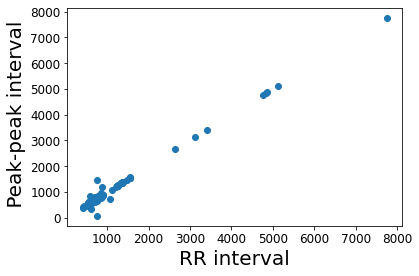}
		\caption{Scatter plot of PPG peak-to-peak interval and ECG RR interval}
		\label{fig:sfig2}
	\end{subfigure}%

	\begin{subfigure}{.4\textwidth}
		\centering
		\includegraphics[width=.8\linewidth]{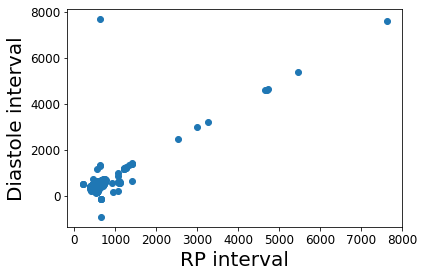}
		\caption{Scatter plots of RP interval of ECG and  diastole period of PPG intervals}
		\label{fig:sfig3}
	\end{subfigure}
	
	\caption{Scatter plots of correlated ECG-PPG intervals}
	\label{fig:scatterplots}

\end{figure}

\textbf{Statistical approach for ECG-PPG correlation:}
 We performed Pearson’s correlation analysis to evaluate the correlation between ECG intervals (RR, PR, RP) and PPG intervals (peak-to-peak interval, systole period, diastole period) for approximately 7586 samples of BIDMC dataset. We plotted the scatter plots for ECG-PPG intervals shown in Figure \ref{fig:scatterplots} and it exhibits linear relationship between the corresponding intervals. Also it is found that the ECG-PPG intervals data follows normal distribution. So, Pearson’s correlation coefficient $r$ is calculated. Significance value ($p$ value) is also computed and $p$ < 0.05 is considered to be significant. The correlational analysis is performed using IBM SPSS version 22 framework \footnote{IBM SPSS Statistics is a software package used for interactive, or batched, statistical analysis \cite{IBMCorp}.} (SPSS Inc, Chicago, IL, USA) \cite{IBMCorp}. 
 
 The Pearson correlation coefficient $r$ and significance value $p$ for RR interval of ECG and peak-to-peak interval of PPG are found to be 0.9941, < .00001 respectively suggesting high degree of correlation between the two ($Obs_1$). The Pearson’s correlation coefficient $r$ and significance values $p$ for other ECG-PPG intervals are presented in Table \ref{table:correlation}. From Table \ref{table:correlation}, it is evident that the PR interval of the ECG is highly correlated to the systole period of the PPG ($Obs_3$), similarly, the RP interval of the ECG is correlated to the diastole period of the PPG ($Obs_4$).

\begin{table}[ht]
	\centering 
	
	\begin{tabular}{|c|l|c|l|c|c|c|}
		\hline
		\multicolumn{2}{|c|}{\multirow{2}{*}{\textbf{\begin{tabular}[c]{@{}c@{}}ECG\\ intervals\end{tabular}}}} & \multicolumn{3}{c|}{\textbf{\begin{tabular}[c]{@{}c@{}}Systole period\\ (onset-systolic peak \\ interval)\end{tabular}}} & \multicolumn{2}{c|}{\textbf{\begin{tabular}[c]{@{}c@{}}Diastole period\\ (systolic peak-onset \\ interval)\end{tabular}}} \\ \cline{3-7} 
		\multicolumn{2}{|c|}{}                                                                                  & \multicolumn{2}{c|}{\textbf{Coef. r}}                               & \textbf{p value}                                 & \textbf{Coef. r}                                           & \textbf{p value}                                           \\ \hline
		\multicolumn{2}{|c|}{PR}                                                                                & \multicolumn{2}{c|}{0.979}                                          & <  0.0001                                & -0.080
		
		                                                     &<0.0001                                          \\ \hline
		\multicolumn{2}{|c|}{QR}                                                                                & \multicolumn{2}{c|}{-0.120}                                         & <  0.0001                                & 0.016                                                     &   0.153                                          \\ \hline
		\multicolumn{2}{|c|}{RP}                                                                                & \multicolumn{2}{c|}{-0.023}                                         & <  0.045                                & 0.788                                                      & <  0.0001                                          \\ \hline
		\multicolumn{2}{|c|}{RT}                                                                                & \multicolumn{2}{c|}{-0.091}                                         & <  0.0001                                & 0.216                                                      & <  0.0001                                          \\ \hline
		\multicolumn{2}{|c|}{QT}                                                                                & \multicolumn{2}{c|}{-0.151}                                         & <  0.0001                                & 0.239                                                     & < 0.0001                                          \\ \hline
	\end{tabular}
	
	\vspace*{-2mm}
	\caption{Statistical analysis of ECG and PPG parameters}
	\label{table:correlation}
	
\end{table}

\textbf{A Formal Runtime Monitoring (RV) Framework for ECG-PPG Correlation:}
Formal \emph{Runtime verification} (RV) \cite{Bauer:2011:RVL,Blech2012,falcone2009runtime,pinisetty2017predictive} 
 a lightweight verification approach dealing with automatic synthesis of RV monitors from a formal high-level specification of the given set of policies.
RV monitors are used to verify a stored execution of a system (offline verification), or the current live execution of a system (online) with respect to a desired correctness policy $\varphi$. 

\begin{figure}[hbt!]
	
	\centering
	
	{
		\includegraphics[width=0.6\linewidth,keepaspectratio]{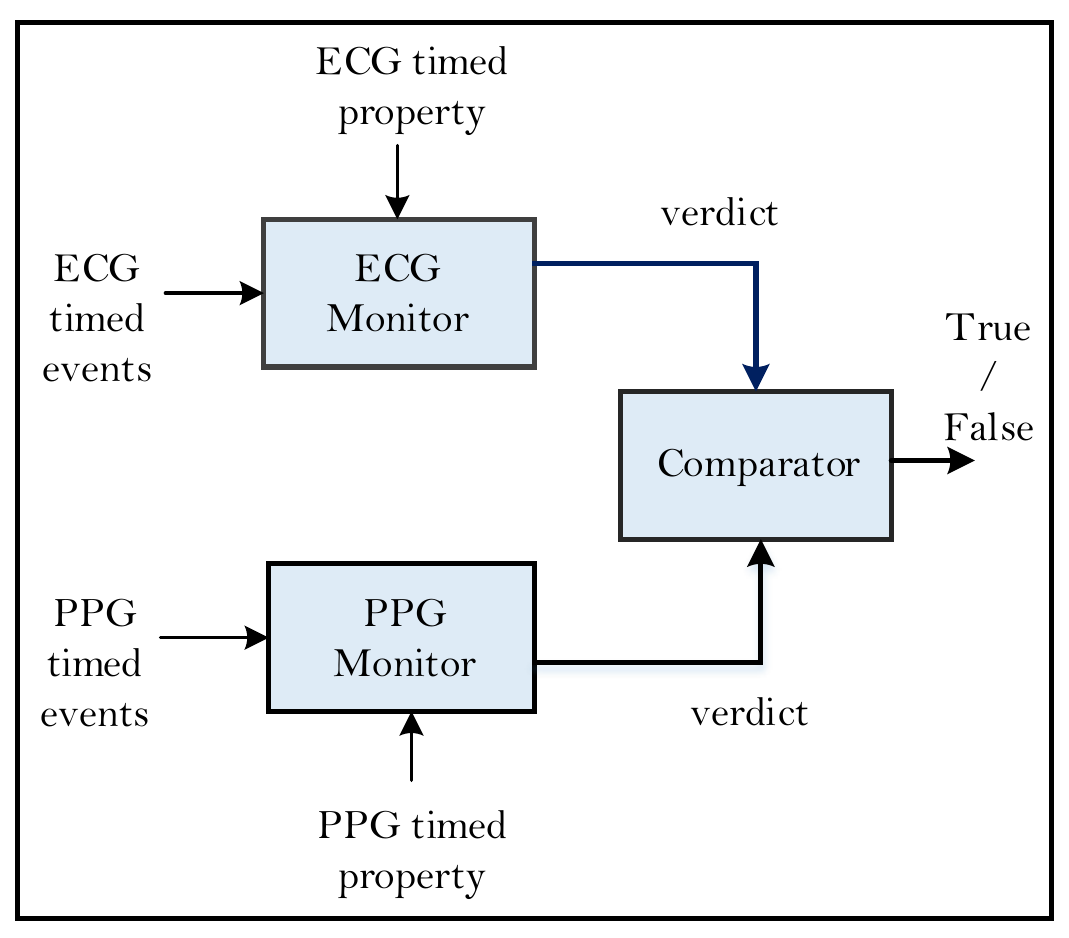}
		\vspace*{-2mm}
		\caption{RV framework for ECG-PPG correlation}
		\label{fig:correlationframework}
	}
	
\end{figure}

 As shown in Figure \ref{fig:correlationframework}, the monitoring framework runs two parallel monitors, ECG and PPG monitor, synthesised from the timed policies (formalized as timed automata (TA)) of the respective signals following the approaches discussed in \cite{pinisetty2017predictive,pinisetty2018security,Bauer:2011:RVL}. 
The events and intervals of ECG-PPG, to be correlated, are input to the respective monitors.  The parallel monitors are fed with traces of the respective  signals. We input the correlated (as per off-line analysis) events of both the signals belonging to the same cardiac cycle and the standard time interval to our monitors to validate the correlation. At each step the verdicts of both the parallel monitors are composed to emit the final verdict. The verdict would be $true$ when the correlation is validated (the verdict $true$ in both the monitors), false otherwise. The following example illustrates the behaviour of the proposed RV framework for the correlations.

\begin{example}[Example illustrating behaviour of the RV framework for correlation]
	Let us consider a scenario of establishing the correlation of the PR interval of ECG to the onset-systolic peak interval (systole period) of PPG. This suggests monitoring the policy: \emph{"The PR interval of ECG and systole period of PPG should be within 210 ms in a cardiac cycle"} (considering the average PR interval of ECG as 210 ms). Consider an ECG trace (events along with time of occurrence): [(p, 728), (r, 888)] where p and r stands for P-peak and R-peak of ECG respectively. The PPG trace corresponding to the ECG cycle is: [(F, 1416), (P, 1568)] where F and P stands for onset and systolic peak of PPG respectively. It may be observed that the events in PPG are delayed, which is due to lag (pulse arrival time) between ECG and PPG events. We fed both the traces to our monitoring frameworks in parallel i.e ECG traces are input to ECG monitor (resp. PPG). We maintained a guard of 210 ms in both the monitoring clocks. This is repeated in a loop throughout the recording. The monitor emits $\false$ whenever there is a violation. The fact that, both the monitors agree in more than 90\% cases, establishes the proposed correlation. Similar approach was adopted for mapping other correlated intervals.
\end{example}

\textbf{Regression analysis of ECG-PPG lag time:}
  We performed linear regression analysis to study the linear relationship ($Obs_2$) and lag time between ECG and PPG events ($Obs_5$, $Obs_6$). We perform regression analysis between correlated ECG-PPG events such as P-peaks of ECG and onsets of PPG, similarly, R-peaks of ECG and systolic peaks of PPG. We implemented a linear regression model based on the regression equation,  $ T_{ppg} = b_1 \times T_{ecg} +  lag\_time $,  where: $T_{ecg}$ and $T_{ppg}$ represents timestamps of correlated ECG event (e.g. R-peak) and PPG event (systolic peak) respectively, $b_1$ denotes slope of the regression line, and $lag\_time$ denotes the lag time between the ECG-PPG event under analysis. The model is implemented in python 3.8 using python package scikit-learn.

\begin{figure}
	\begin{subfigure}{.25\textwidth}
		\centering
		\includegraphics[width=.8\linewidth]{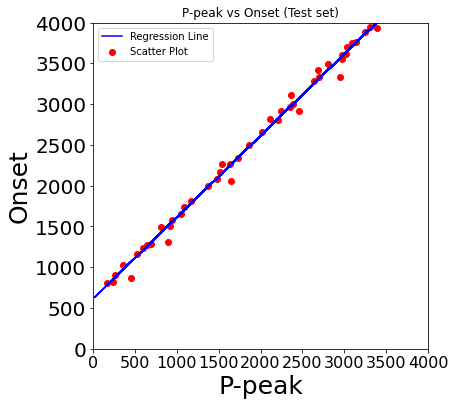}
		\caption{Regression analysis of \\P-peaks of ECG  and onsets\\ of PPG}
		\label{fig:sfig4}
	\end{subfigure}%
	\begin{subfigure}{.25\textwidth}
		\centering
		\includegraphics[width=.8\linewidth]{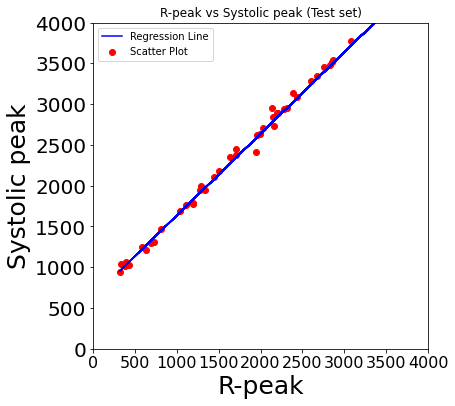}
		\caption{Regression analysis of R-peaks of ECG and systolic peaks of PPG}
		\label{fig:sfig5}
	\end{subfigure}

	\caption{Regression analysis of correlated ECG-PPG events}
	\label{fig:regression}
	
\end{figure}

Figure \ref{fig:sfig4} represents the linear regression between P-peak of ECG and onset of PPG. The estimated coefficients are found to be 0.999 ($b_1$) and 607.065 ms ($lag\_time$) for the regression analysis. The value of the coefficient,  $b_1$ = 0.999, suggest the linear relationship between the two events ($Obs_2$) and the value of the coefficient,  $lag\_time$= 607.065 ms suggest the lag time to be in the range of 600-700 ms validating the analysis ($Obs_5$). Similarly, Figure \ref{fig:sfig5} represents the linear regression between R-peak of ECG and systolic peak of PPG. The coefficients are found to be 0.999 ($b_1$) and 606.486 ms ($lag\_time$) for the regression analysis. The value of the coefficient,  $b_1$ = 0.999, suggest the linear relationship between the two events and the value of the coefficient,  $lag\_time$= 606.486 ms, suggest the lag time to be in the range of 600-700 ms validating the observation ($Obs_6$).

So, using the above analysis, it is concluded that observations $Obs_1$:The RR interval of ECG is correlated to the peak-to-peak interval of PPG, $Obs_2$: The P-peak of ECG corresponds to the onset of PPG signal, $Obs_3$: The PR interval of ECG is correlated to the systole period of PPG, $Obs_4$: The RP interval of ECG is correlated to the diastole period of PPG, $Obs_5$: The onset of PPG lags the P-peak of ECG by 600 - 700 ms and $Obs_6$: The systolic peak of PPG lags  the R-peak of ECG by 600 - 700 ms  are valid and hence represent the correlation between ECG and PPG events and intervals.
 

\section{Conclusions}
\label{sec:conclusion}
In this paper, we propose the correlation of ECG-PPG events and intervals and validate the correlation using formal monitoring and statistical approaches. The proposed analysis and monitoring framework can be tested using lager sample size/data sets. Such correlations will provide better insight for understanding the health parameters of an individual.  The synthesized runtime monitors used for validating the correlation  would also be useful in continuous health monitoring applications and detecting attacks and securing implantable medical devices using ECG and PPG.


\bibliographystyle{IEEEtran}
\bibliography{biblioShort}

\end{document}